\documentclass{elsart}
\def\maru#1{\textcircled{\scriptsize#1}}
\usepackage{graphicx}
\usepackage{amssymb}
\begin{document}
\begin{frontmatter}
\title{Scintillation yield of liquid xenon at room temperature}
\author{K.Ueshima\thanksref{label1}\corauthref{cor1}}, 
\ead{ueshima@suketto.icrr.u-tokyo.ac.jp}
\corauth[cor1]{Corresponding Author}
\author{K.Abe\thanksref{label1}},
\author{T.Iida\thanksref{label1}},
\author{M.Ikeda\thanksref{label1}},
\author{K.Kobayashi\thanksref{label1}},
\author{Y.Koshio\thanksref{label1}},
\author{A.Minamino\thanksref{label1}},
\author{M.Miura\thanksref{label1}},
\author{S.Moriyama\thanksref{label1}},
\author{M.Nakahata\thanksref{label1}\thanksref{label22}},
\author{Y.Nakajima\thanksref{label1}},
\author{H.Ogawa\thanksref{label1}},
\author{H.Sekiya\thanksref{label1}},
\author{M.Shiozawa\thanksref{label1}},
\author{Y.Suzuki\thanksref{label1}\thanksref{label22}},
\author{A.Takeda\thanksref{label1}},
\author{Y.Takeuchi\thanksref{label1}},
\author{M.Yamashita\thanksref{label1}},
\author{K.Kaneyuki\thanksref{label11}},
\author{T.Doke\thanksref{label2}},
\author{Y.Ebizuka\thanksref{label2}},
\author{J.Kikuchi\thanksref{label2}},
\author{A.Ota\thanksref{label2}},
\author{S.Suzuki\thanksref{label2}},
\author{T.Takahashi\thanksref{label2}},
\author{H.Hagiwara\thanksref{label3}},
\author{T.Kamei\thanksref{label3}},
\author{K.Miyamoto\thanksref{label3}},
\author{T.Nagase\thanksref{label3}},
\author{S.Nakamura\thanksref{label3}},
\author{Y.Ozaki\thanksref{label3}},
\author{T.Sato\thanksref{label3}},
\author{Y.Fukuda\thanksref{label4}},
\author{T.Sato\thanksref{label4}},
\author{K.Nishijima\thanksref{label5}},
\author{M.Sakurai\thanksref{label5}},
\author{T.Maruyama\thanksref{label5}},
\author{D.Motoki\thanksref{label5}},
\author{Y.Itow\thanksref{label6}},
\author{H.Ohsumi\thanksref{label7}},
\author{S.Tasaka\thanksref{label8}},
\author{S.B.Kim\thanksref{label9}},
\author{Y.D.Kim\thanksref{label10}},
\author{J.I.Lee\thanksref{label10}},
and
\author{S.H.Moon\thanksref{label10}}\\
{the XMASS Collaboration}
\address[label1]{Kamioka Observatory, Institute for Cosmic Ray Research, The University of Tokyo, Kamioka, Hida, Gifu 506-1205, Japan}
\address[label11]{Research Center for Cosmic Neutrinos, Institute for Cosmic Ray Research, The University of Tokyo, Kashiwa, Chiba 277-8582, Japan}
\address[label22]{Institute for the Physics and Mathematics of the Universe, The University of Tokyo, Kashiwa, Chiba 277-8582, Japan}
\address[label2]{Faculty of Science and Engineering, Waseda University, Shinjuku-ku, Tokyo 162-8555, Japan }
\address[label3]{Department of Physics Faculty of Engineering, Yokohama National University, Yokohama, Kanagawa 240-8501, Japan}
\address[label4]{Department of Physics, Miyagi University of Education, Sendai, Miyagi 980-0845, Japan}
\address[label5]{Department of Physics, Tokai University, Hiratsuka, Kanagawa 259-1292, Japan}
\address[label6]{Solar Terrestrial Environment Laboratory, Nagoya University, Nagoya, Aichi 464-8602, Japan }
\address[label7]{Faculty of Culture and Education, Saga University, Honjo, Saga 840-8502, Japan}
\address[label8]{Department of Physics, Gifu University, Gifu, Gifu 501-1193, Japan}
\address[label9]{Department of Physics, Seoul National University, Seoul 151-742, Korea}
\address[label10]{Department of Physics, Sejong University, Seoul 143-747, Korea}

\begin{abstract}
The intensity of scintillation light emission from liquid xenon at room temperature was measured.
The scintillation light yield at 1 $^{\circ}$C was measured to be 0.64 $\pm$ 0.02 (stat.) $\pm$ 0.06 (sys.) of that at -100$^{\circ}$C.
Using the reported light yield at -100 $^{\circ}$C (46 photons/keV), the measured light yield at 1 $^{\circ}$C
corresponds to 29 photons/keV.
This result shows that liquid xenon scintillator gives high light yield even at room temperature.
\end{abstract}

\end{frontmatter}

\section{Introduction}
\begin{figure}[p]
\begin{center}
\includegraphics[width=8cm,clip]{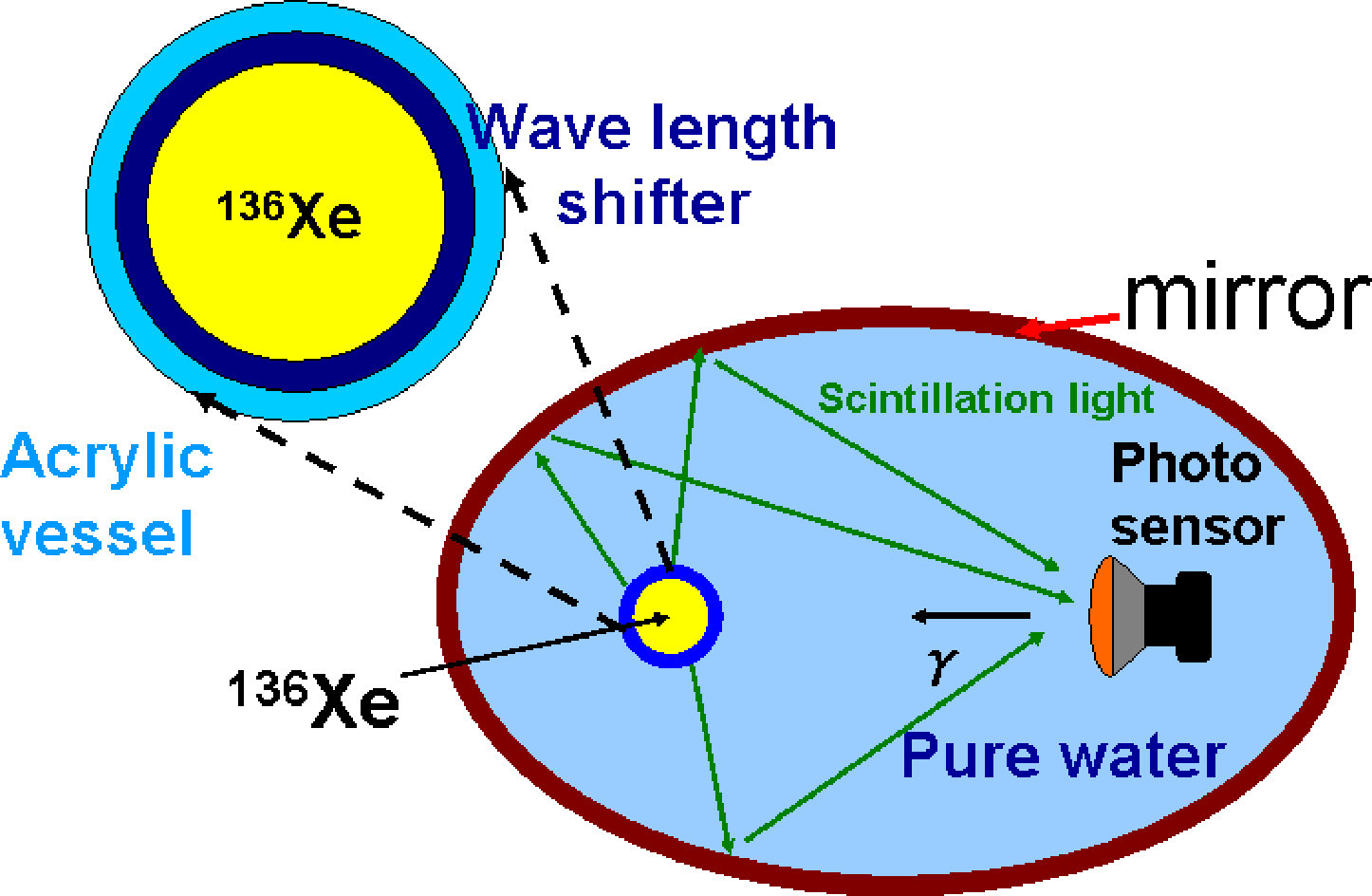}
\caption{Conceptual design of a double beta decay experiment. The ELT detector consists of photosensors and enriched liquid xenon. Since the inner surface of ELT is mirrored, 
the scintillation light from $^{136}$Xe is detected with high efficiency. Furthermore, the photosensors which 
are the main BG source are kept away
from $^{136}$Xe.}
\label{fig:elt}
\end{center}
\end{figure}

Scintillation detectors based on liquid xenon have been constructed for various experiments, such as 
dark matter searches [1], double beta decay [2], $\mu\rightarrow$e$\gamma$ search [3], etc. 
The great advantages of liquid xenon are its large scintillation yield,
which enables us to measure low energy phenomena and perform experiments which require good energy resolution.
The light yield of liquid xenon at -100$^{\circ}$C has been measured by many experiments [4-9].
The scintillation light of liquid xenon is vacuum ultraviolet(VUV). The mechanism of light emission was reported in [10]. 
Because of low boiling point of xenon (-100$^{\circ}$C at 0.18 MPa absolute pressure ), a cooling medium or a refrigerator is necessary to make detectors 
using liquid xenon.
If it is possible to use liquid xenon at room temperature, it would make detector construction easier and enable various new possibilities 
for detectors.
The density of liquid xenon at room temperature is 65\% of that at -100$^{\circ}$C (1.88 g/c$m^3$ at 1$^{\circ}$C) [11], but 
it is still a high density scintillator.
Xenon can be in liquid phase below the critical point of 17$^{\circ}$C.\\
As an example of new applications by using liquid xenon at room temperature, a detector for double beta decay which requires a low background technique is discussed here.
The conceptual design of the detector is shown in Fig.\ref{fig:elt}.
The detector consists of an elliptic tank (ELT) filled with water, an acrylic vessel filled with enriched $^{136}$Xe and a few photosensors.
The acrylic vessel is put at one focus and photosensors are put at the other focus. Since the scintillation light of liquid xenon 
is VUV, it is not possible to transmit through the acrylic vessel and water, threfore wavelength shifter is necessary on the inner surface of 
the acrylic vessel.
The inner surface of the ELT is a reflector for the wavelength-shifted photons with high efficiency in order to guide photons to the 
photosensors. 
Because of the convergence of photons to a relatively small focus point, we can dramatically reduce the number of photosensors. 
Usually, one of the most serious sources of background in low background experiments is radioactivity in the photosensors.
The large reduction of the number of photosensors helps a
lot in reducing the background.
Furthermore,
because the gamma rays from the photosensors 
 are attenuated by pure water in the ELT, it is possible to reduce the backgrounds further.\\ 

  The scintillation yield of liquid xenon at -100 $^{\circ}$C has been reported in several papers [12], but the yield at 
room temperature has not been reported so far.
In this paper, we present the measurement of scintillation light yield at room temperature and compare to that at -100 $^{\circ}$C.

\section{Experimental setup}
\subsection{Liquid xenon vessel}
\begin{figure}[p]
\begin{center}
\includegraphics[width=8cm,clip]{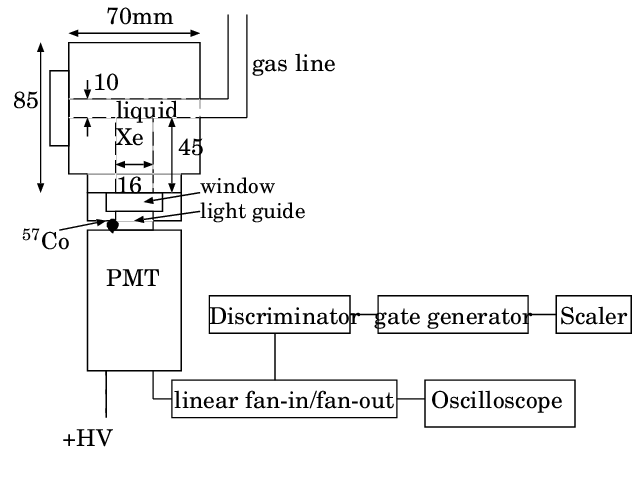}
\caption{Cross section of the liquid xenon vessel and measurement diagram; the inner volume of the high pressure vessel 
is 15 cc. The $^{57}$Co source is located at edge of light guide.}
\label{fig:cell}
\end{center}
\end{figure}

 Figure \ref{fig:cell} shows the cross section of the stainless-steel liquid xenon vessel (SUS304). The vertical cylindrical hole (16 mm diameter, 45 mm length) and the horizontal cylindrical hole (10mm diameter, 70 mm length)  
 is completely filled with liquid xenon and its volume is 15 cc. The pressure of liquid xenon at room temperature is 5 MPa. 
 Therefore, the body of the vessel is made from stainless steel with a thickness of 27 mm and the window where scintillation light passes
 is made from MgF$_2$ with a thickness of 10 mm.
 In order to transmit scintillation light to a photomultiplier (PMT), a light guide made from MgF$_2$ is placed between the 
 MgF$_2$ window and the PMT. The light guide is needed because oxygen in air easily absorbs the VUV scintillation light.
 The diameter of the light guide is 16 mm and the length is 10 mm. A 2-inch PMT (R8778 Hamamatsu), which is sensitive to VUV light,
  is used in this measurement.
 The gain of the PMT is set to 6.0$\times10^6$ with an applied high voltage of +1544V. The quantum
  efficiency(Q.E.) of the PMT is 28\% at room temperature for 175 nm VUV light. In order to compare measurements at different temperature, the temperature dependence of the Q.E. and gain were measured independently. As a results, it was revealed that the product of them, Q.E. $\times$ gain increased 15$\pm$5 \% from 1$^{\circ}$C to -100$^{\circ}$C.
The temperature dependence of the gain and Q.E. are taken into account in the following measurements.
In order to collect scintillation light efficiently, Dupont$^{\maru{TM}}$ Krytox$^{\maru{R}}$ 16350 is used as an optical grease on the both side of the light guide.
We confirmed that the transmittance of the Krytox at 175 nm was more than 90\% with the thickness of 25 $\mu$m.
 The refractive index of Krytox is estimated to be 1.35 at 175 nm.
 
\subsection{Filling procedure} 
\begin{figure}[p]
\begin{center}
\includegraphics[width=10cm,clip]{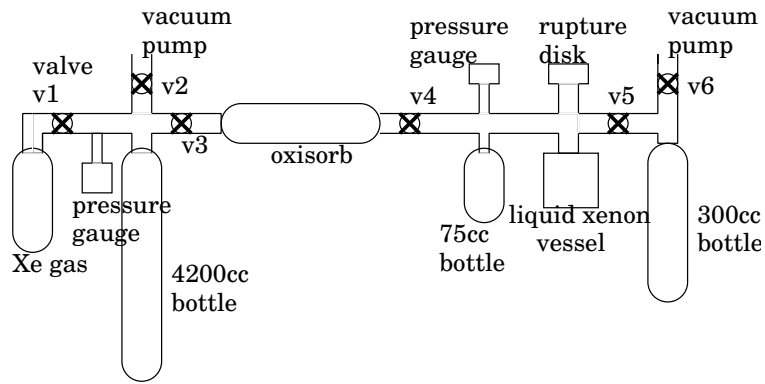}
\caption{
Handling system of high pressure gas
to fill liquid xenon at room temperature into the liquid xenon vessel.
Oxisorb reduces impurities in xenon gas.}
\label{fig:gas}
\end{center}
\end{figure}

Figure \ref{fig:gas} shows a handling system
of high pressure gas to fill liquid xenon at room temperature into the liquid xenon vessel.
First, xenon gas was collected in the 4200 cc bottle. The pressure of xenon gas in the 4200 cc bottle was 0.5 MPa,
which corresponds to 110 $g$ of xenon in the bottle. Second, the collected xenon gas was passed through the Oxisorb (MESSER; large
cartridge Oxisorb).
The Oxisorb filters out oxygen and water from xenon gas to the level of oxygen ƒ 5ppb and H$_2$O ƒ 30ppb.
In order to fill xenon into the vessel using only temperature cycle (i.e., without using a compressor), 
a 75 cc bottle was placed between Oxisorb and the vessel.
 The 75 cc bottle was cooled down in advance using liquid nitrogen.  Accordingly, the filtered xenon gas 
 was transferred to the 75 cc bottle. Finally, by raising the temperature of
 the 75 cc bottle, the liquid xenon is transferred to the vessel. 
 The pressure of liquid xenon was measured using a pressure gauge (KYOWA PHS-200KA). 
 It was kept at 5.8$\pm$0.1 MPa. The temperature of the liquid xenon vessel was kept at 1 $\pm$ 2$^{\circ}$C
 using a coolant placed on the vessel. 
 The 75 cc bottle was connected to the liquid xenon vessel throughout the measurement in order to keep the vessel
 always filled with liquified xenon. 
 For safety, a rupture disk which bursts at 13 MPa was connected to the liquid xenon vessel.  

\subsection{Setup for the reference measurement}
\begin{figure}[p]
\begin{center}
\includegraphics[width=11cm,clip]{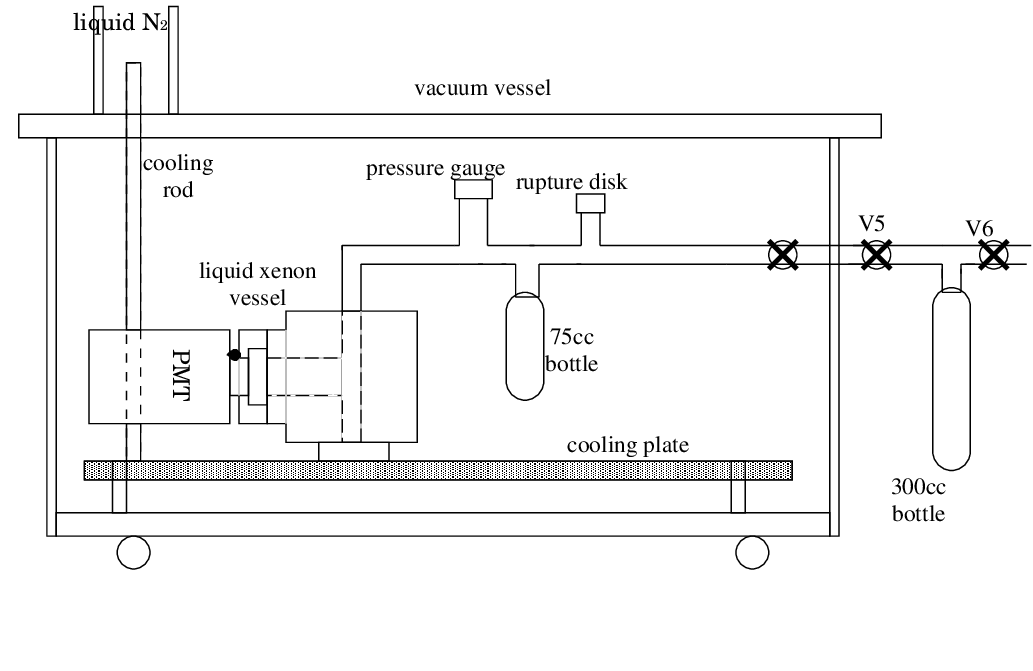}
\caption{Setup for the reference measurement; the liquid xenon vessel is put in a vacuum chamber and is cooled down using liquid nitrogen.
The 300 cc bottle is used to drain xenon gas from the liquid xenon vessel.}
\label{fig:teicell}
\end{center}
\end{figure}

 We measured relative light yield at room temperature with respect to that at reference temperature (T=-100$^{\circ}$C,P=0.18 MPa). 
 Figure \ref{fig:teicell} shows the setup for the reference measurement.
 The liquid xenon vessel, together with the 75 cc bottle and the PMT, is placed in a vacuum chamber.
 There is a liquid nitrogen holder on the vacuum chamber and it is connected to a
 cooling rod and cooling plate in the vacuum chamber. The liquid xenon vessel is cooled down by the cooling plate.  
 The temperature was controlled using a heater wrapped around the cooling rod; the temperature of the liquid xenon vessel is kept at
 -100$\pm$2$^{\circ}$C.
 In order to control the pressure in the vessel, a 300 cc bottle 
 is connected to the gas line as shown Fig.\ref{fig:teicell}. 

\section{Measurement}
  The scintillation light of liquid xenon was measured using 122 keV gamma rays emitted from a $^{57}$Co source with 7.4 kBq.
The $^{57}$Co source was put at the edge of the light guide as shown in Fig.\ref{fig:cell}.
In liquid xenon, the 122 keV gamma rays are absorbed within few mm. 
 Hence, most of the gamma rays caused scintillation light emission near the window.
 The signal from the PMT was measured by a digital oscilloscope (TDS 3064B). 
 The trigger threshold was set to -16 mV (3 photoelectron level) and the pulse waveform was recorded with 500MHz sampling rate.
 Data acquisition time for each event is about one second. In order to measure the event rate, the PMT signal was divided by a
 linear fan-in$\slash$fan-out to the oscilloscope
 input and to a NIM discriminator with a threshold of -16 mV. The output from the discriminator was connected to a scaler.\\
 First, the scintillation light at room temperature was measured.
 Since the critical point of xenon is 17$^{\circ}$C, the temperature of the liquid xenon was kept at
 1$\pm$2 $^{\circ}$C which was well below the critical point. The pressure of the liquid xenon was 5.5 MPa. 
 The event rate with the $^{57}$Co source was 80 Hz, while the event rate of background (i.e., without source) was 57 Hz.\\
 Secondly, the scintillation light at -100 $^{\circ}$C was measured. 
The liquid xenon vessel was set on the cooling plate in the vacuum chamber.
 Using the liquid nitrogen and the cooling plate, the liquid xenon vessel was cooled down from room temperature to -100 $^{\circ}$C. During cooling of the vessel, the temperature dependence of scintillation yield was measured.\\
 Since the liquid xenon vessel was connected to the 75 cc bottle whose temperature was still room temperature, the
 pressure of liquid xenon was still high around 4.5 MPa even at -100$^{\circ}$C.
 In order to compare the scintillation yield at (T=-100 $^{\circ}$C, P=0.18 MPa) with the scintillation yield at (T=-1 $^{\circ}$C, P=5.5 MPa),
  the pressure of liquid xenon had to be reduced from 4.5 MPa to 0.18 MPa at -100 $^{\circ}$C. 
 Xenon gas was drained from the liquid xenon vessel to the 300 cc bottle as shown in Fig.\ref{fig:teicell}. When the pressure in
 the 300 cc bottle was about 0.4 MPa, 
 xenon gas was drained from the 300 cc bottle to atmosphere by opening V6 (V5 was closed before hand).
 After draining, the 300 cc bottle was evacuated. 
 We continued to drain xenon gas
 from the liquid xenon vessel to atmosphere until the pressure in the liquid xenon vessel was reduced to 0.18 MPa at -100 $^{\circ}$C.
 Even during draining, 
 the temperature of the liquid xenon vessel was kept to be -100 $\pm$ 2 $^{\circ}$C.
 Scintillation light yield was measured at each step of draining xenon, thus pressure dependence was measured.
 Finally, the scintillation light at (T=-100$^{\circ}$C, P=0.18 MPa) was measured. The event rate at this temperature
 was 61 Hz, while the BG rate was 7.1 Hz.
 Since the PMT dark rate decreased at low temperature, the event rate was decreased.
  
\section{Data analysis}
\subsection{Selecting scintillation light events}
\begin{figure}[p]
\begin{center}
\includegraphics[width=7cm,clip]{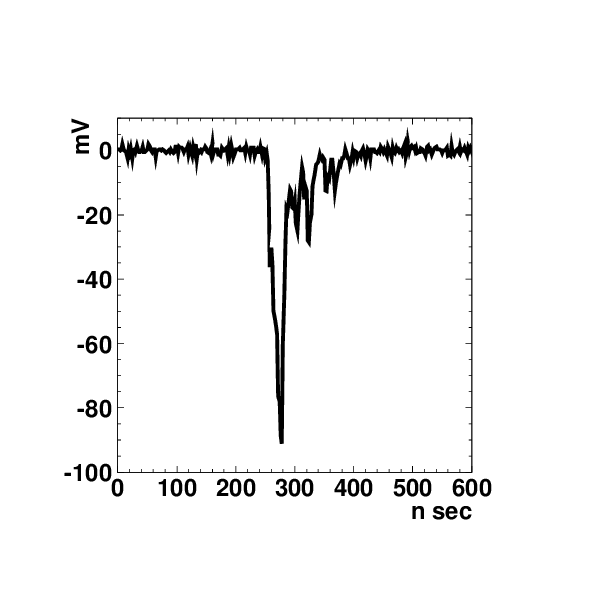}
\caption{A typical waveform of a scintillation event. The observed charge corresponds 36p.e.. }
\label{fig:1}
\end{center}
\end{figure}

\begin{figure}[p]
\begin{center}
\includegraphics[width=7cm,clip]{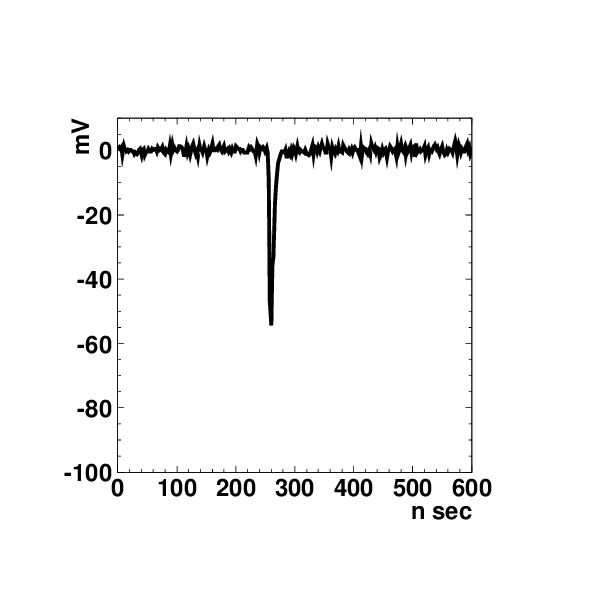}
\caption{A typical waveform of a PMT dark noise event. The observed charge corresponds to 6p.e..}
\label{fig:2}
\end{center}
\end{figure}

\begin{figure}[p]
\begin{center}
\includegraphics[width=7cm,clip]{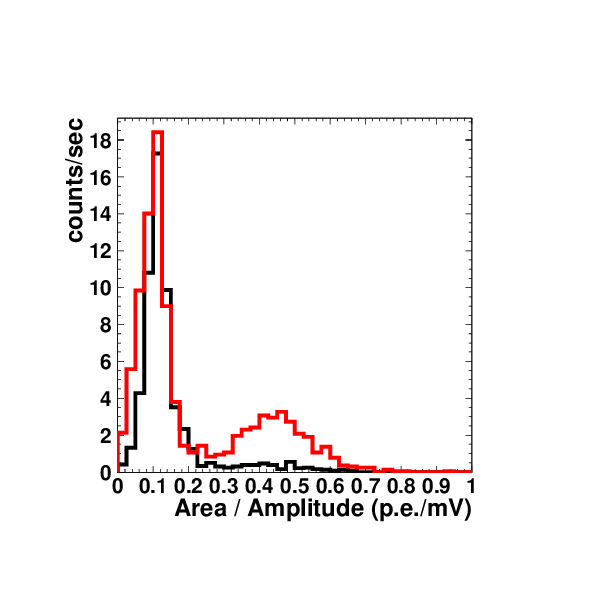}
\caption{ \textit{area/amplitude} (p.e./mV) distribution for $^{57}Co$ source data (red) and BG data (black). The peaks at \textit{area/amplitude} $\sim$ 0.1
and $\sim$ 0.45 are due to PMT dark noise and scintillation signals, respectively.}
\label{fig:de}
\end{center}
\end{figure}

\begin{figure}[p]
\begin{center}
\includegraphics[width=7cm,clip]{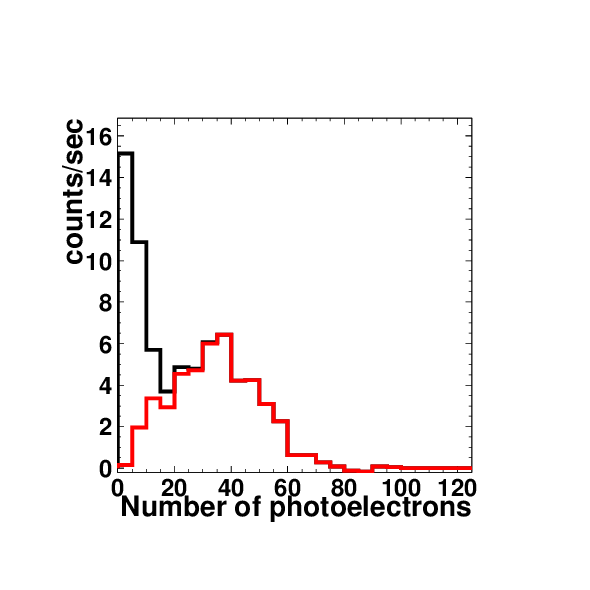}
\caption{ pulse height distribution of $^{57}Co$ source data; 
The black and red lines show the p.e. distribution before and after the \textit{area/amplitude} cut is applied, respectively.}
\label{fig:cut}
\end{center}
\end{figure}

 The majority of the BG events are due to PMT dark noise. These events usually have a short pulse width compared with the scintillation signal of
 liquid xenon.
 The decay time constant of scintillation light of liquid xenon is about 45 ns [13]. 
 Figure \ref{fig:1} shows a pulse waveform of a typical scintillation signal, and Fig.\ref{fig:2} shows that of a dark noise signal.
 In order to discriminate scintillation light from dark noise,
 a cut using pulse shape was applied. In order to evaluate the width of the pulse, the peak amplitude of the signal ($\textit{pulse amplitude}$)
  and the
 integrated area of the signal ($\textit{pulse area}$) are used.
 The $\textit{pulse area}$ is the amount of integrated charge for every 2 ns with a total integration time window of 300 ns. 
 Figure \ref{fig:de} shows the ratio of $\textit{pulse area}$ and $\textit{pulse amplitude}$ (\textit{area/amplitude}).
 The black line shows the ratio for BG data and the red line shows $^{57}$Co source data.  
 There are two peaks in the source data. 
 The peak at \textit{area/amplitude} of $\sim$ 0.1 p.e./mV is due to sharp pulse events
 caused by dark noise of the PMT.   
 On the other hand, the peak at \textit{area/amplitude} of $\sim$ 0.45 p.e./mV is due to wide pulse events caused by scintillation 
 light.
 To choose scintillation events, a cut criterion of \textit{area/amplitude} $>$ 0.19 was applied.
 In order to make a pulse height distribution of pure $^{57}$Co gamma ray events, the \textit{area/amplitude} cut and BG subtraction were 
 applied.
 Figure \ref{fig:cut} shows the pulse height distributions with the BG subtraction before and after the \textit{area/amplitude} cut. 
 The event rate due to 122 keV gamma rays was obtained by subtracting the BG rate from the data with the source and applied the \textit{area/amplitude} cut.
 The event rate of 122 keV gamma rays was 53Hz at 1$^{\circ}$C and 56Hz at -100$^{\circ}$C. No significant change in \textit{area/amplitude} distribution of gamma ray signals was observed over temperature and pressure variations.  

\subsection{Observed light yield, pressure and temperature dependence}
\begin{figure}[p]
\begin{center}
\includegraphics[width=7cm,clip]{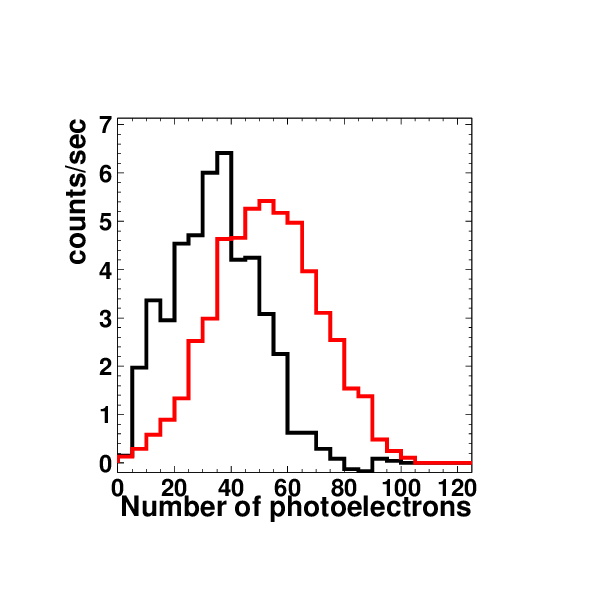}
\caption{p.e. distribution at (T=1 $^{\circ}$C, P=5.5 MPa). The black line shows the data. The red line shows the MC.}
\label{fig:high}
\end{center}
\end{figure}

\begin{figure}[p]
\begin{center}
\includegraphics[width=7cm,clip]{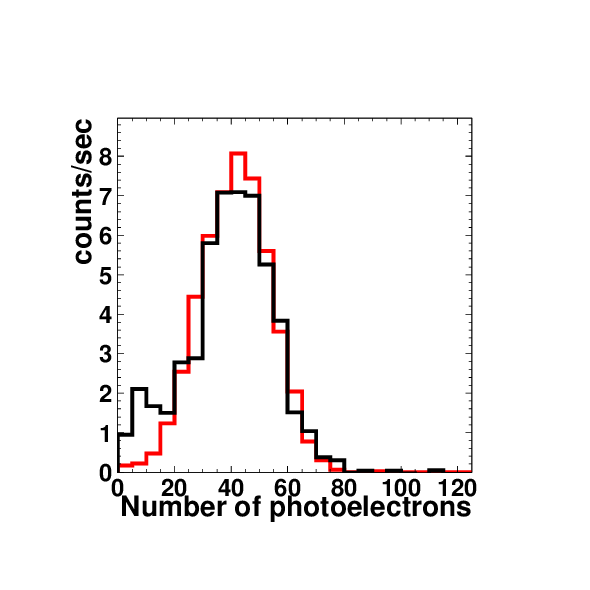}
\caption{p.e. distribution at (T=-100 $^{\circ}$C, P=0.18 MPa).  The black line shows the data. The red line shows the MC.}
\label{fig:low}
\end{center}
\end{figure}

\begin{figure}[p]
\begin{center}
\includegraphics[width=7cm,clip]{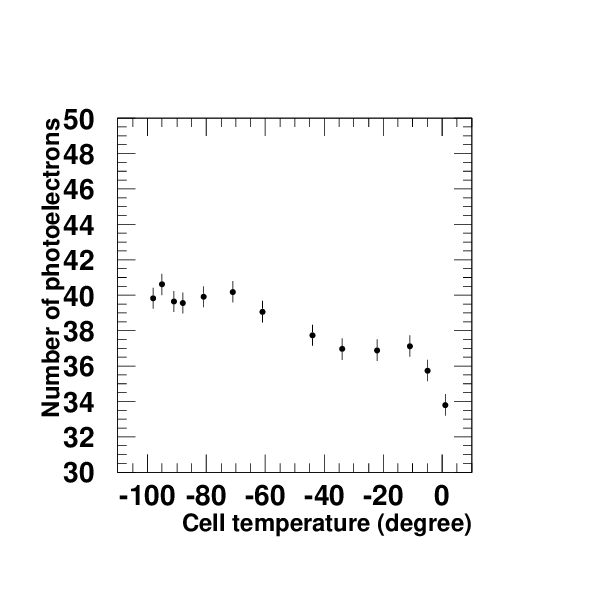}
\caption{Measured temperature dependence of the number of photoelectrons. The measured pulse height decreased by
19 \% from -100 $^{\circ}$C to 1 $^{\circ}$C. The vertical error is statistical error.}
\label{fig:temp}
\end{center}
\end{figure}

\begin{figure}[p]
\begin{center}
\includegraphics[width=7cm,clip]{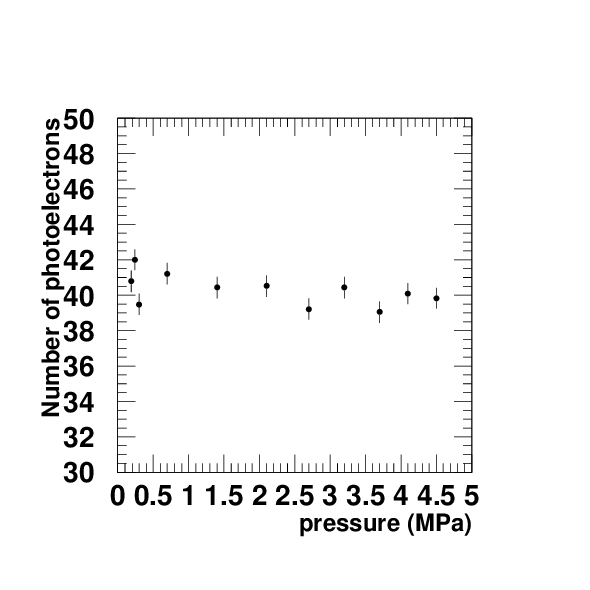}
\caption{pressure dependence of the pulse height measured at T = -100$^{\circ}$C. It is stable within 4\%. The vertical error is statistical error.}
\label{fig:pres}
\end{center}
\end{figure}

 The black line in Fig.\ref{fig:high} shows the pulse height distribution of liquid xenon at (T=1 $^{\circ}$C, P=5.5 MPa) 
 after the \textit{area/amplitude} cut is applied.
 The mean value with a Gaussian fit is 33.8$\pm$0.6 p.e.
 On the other hand, the black line in Fig.\ref{fig:low} shows the pulse height distribution of liquid xenon at (T=-100 $^{\circ}$C, P=0.18 MPa). 
 The mean value for this case is 41.7$\pm$0.6 p.e. As a result of the \textit{area/amplitude} analysis, the shape of the pulse height distribution at
  (T=1 $^{\circ}$C, P=5.5 MPa) and (T=-100 $^{\circ}$C, P=0.18 MPa) is consistent within error.
 Figure \ref{fig:temp} shows the temperature dependence of 
 the measured light yield. The number of observed photoelectrons (p.e.) increases at low temperature and the difference in light yield between 
 at -100 $^{\circ}$C and at 1 $^{\circ}$C is 19\%. 
 In order to obtain the relative scintillation light yield between at -100 $^{\circ}$C and at 1 $^{\circ}$C, the difference in refractive 
 index between 
 those temperatures must be taken into account. This is because the acceptance of the PMT depends on the refractive index of the liquid xenon
  (described in more detail in the next section).
 Figure \ref{fig:pres} shows the pressure dependence from 4.5 MPa to 0.18 MPa at -100 $^{\circ}$C.
 The number of p.e. is stable within 4\%. 

\subsection{Comparison of the light yield}
A comparison  of the scintillation light yield between -100 $^{\circ}$C and 1 $^{\circ}$C is discussed in this section.
As described in the previous section, in order to obtain the relative scintillation light yield from the measured pulse height, 
the refractive index of liquid xenon must be taken into account.
The density ($\rho$, g/c$m^3$) at (T=-100 $^{\circ}$C, P=0.18 MPa) and (T=1 $^{\circ}$C, P=5.5 MPa) is 2.88 g/c$m^3$ and 1.88 g/c$m^3$, respectively. 
Generally, the refractive index changes if the material density changes. 
The following equation shows the refractive index as a function of density [14]: 
\begin{equation}
n = \sqrt{\frac{(1+2x)n_0^2 + (2-2x)}{(1-x)n_0^2 + (2+x)}}                    
\end{equation}
where \textit{n$_0$} (=1.62) [15] is the refractive index at (T=-100 $^{\circ}$C, P=0.18 MPa) and \textit{x} is the density ratio, which is
$\rho$(1$^{\circ}$C, 5.5 MPa) / $\rho$(-100$^{\circ}$C, 
0.18 MPa). In this case \textit{x} is 0.65. 
From this equation the refractive index at (T=1 $^{\circ}$C, P=5.5 MPa) is 1.37.
Because the refractive index of the MgF$_2$ window is 1.44 [16], the acceptance of the PMT increases as
the refractive index of the liquid xenon becomes smaller.
Hence, the acceptance for (T=1$^{\circ}$C, P=5.5 MPa) is larger than that for (T=-100 $^{\circ}$C, P=0.18 MPa).  
In order to estimate the difference of the acceptance, a simulation based on the GEANT4 [17] package was used. Table \ref{tab1} shows 
input physical values to the simulation. 
The light yield used in the simulation (51 photons/keV) is 10 \% higher than the nominal value of the light yield (46 photons/keV).
 This is because the nominal value is given for 1 MeV electrons, while the
value for 122 keV gamma ray must be used in our simulation. The energy dependence of the light yield for electrons and gamma ray are taken from ref.[18].
The ${\it tuning}$ ${\it factor}$ in the table \ref{tab1} is the absolute normalization factor to reproduce the observed charge at (T=-100 $^{\circ}C$, P=0.18 MPa) by the MC simulation for the corresponding condition.
\begin{table}[h]
\begin{center}
\caption{Physical values}
\begin{tabular}{|c|c|} \hline 
light yield at (T=-100 $^{\circ}$C, P=0.18 MPa)& 51 photons/keV \\ \hline
density & 1.88 g/c$m^3$(1$^{\circ}$C), 2.88 g/c$m^3$(-100$^{\circ}$C)\\ \hline
refractive index & 1.37(1$^{\circ}$C),1.62(-100$^{\circ}$C) [15] \\ \hline
PMT Q.E. & 28\% \\ \hline
MgF$_2$ refractive index & 1.44 \\ \hline
MgF$_2$ absorption length & 14.6 cm \\ \hline
tuning factor & 0.84 \\ \hline
\end{tabular}
\label{tab1}
\end{center}
\end{table}
Using same physical values except for density and refractive index,  
the difference of the p.e. distribution between (T=-100 $^{\circ}$C, P=0.18 MPa) and (T=1 $^{\circ}$C, P=5.5 MPa) was simulated.
The red line in Fig.$\ref{fig:high}$ and Fig.\ref{fig:low} shows the result
 of the simulation.  At (T=-100 $^{\circ}$C, P=0.18 MPa) the mean value of pulse height (M$_{ph}$) is 41.5$\pm$0.3 p.e.
 The simulated M$_{ph}$ for (T=1 $^{\circ}$C, P=5.5 MPa) is 52.7$\pm$0.4 p.e.,
 while the observed M$_{ph}$ for this condition was 33.8$\pm$0.6 p.e..
 The relative scintillation light yield was obtained by the following equation : 
\begin{equation}
eff_{acc} = \frac{M_{ph}(-100^{\circ}C,MC)}{ M_{ph}(1^{\circ}C,MC)}
\end{equation}
\begin{equation}
Ratio = \frac{M_{ph}(1^{\circ}C,data)}{M_{ph}(-100^{\circ}C,data)} \times eff_{acc}
\end{equation}
 As a result, the ratio of scintillation yield is
  \begin{equation}
Ratio = 0.64 \pm 0.02 (stat.) \pm 0.06 (sys.)
\end{equation}
The systematic error includes reproducibility error (3\%), 
the increase of the PMT Q.E.$\times$gain error (5\%).
\section{Summary and Discussion}
The scintillation light yield of liquid xenon at room temperature was measured.
In this work, the scintillation light yield at room temperature (T=1$^{\circ}$C,P=5.5 MPa) is 0.64$\pm$0.02$\pm$0.06
 of that at usual liquid xenon temperature (T=-100$^{\circ}$C,P=0.18 MPa). Using the reported scintillation light yield at usual
 liquid xenon temperature (46 photons/keV), the light yield at room temperature corresponds to 29 photons/keV.
 Pressure dependence of the light yield at T = -100$^{\circ}$C was also measured, and it is less than 4 \% between 0.18 MPa and 4.5 MPa.\\
 The obtained scintillation light yield is high, giving a possibility for the construction of new particle detectors.
 For example, a double beta decay experiment, as shown in Fig.1, is possible.
 Double beta decay experiments require high energy resolution, because 2$\nu\beta\beta$ must be rejected from 0$\nu\beta\beta$
 signal region. Assuming the following conditions in the experimental setup of Fig.1,
\begin{itemize}
\item Efficiency of wavelength shifter : 50 \% 
\item Photon collection efficiency of ELT : 50 \%, 
\item Quantum efficiency of photo-detector : 25 \%, 
\end{itemize} 
the number of expected photoelectrons at the Q value of 0$\nu\beta\beta$ of $^{136}$Xe (2.47 MeV) is 4000 p.e.. 
If the energy resolution
is determined only by photoelectron
statistics, the detector
gives 1.6\%$_{RMS}$ in the energy resolution. However, this
needs to be studied experimentally [19]. 
It should enable us to test the 
 lifetime of 0$\nu\beta\beta$ of $^{136}$Xe up to $\sim$ $10^{27}$years.
 
\section{Acknowledgements}
 We gratefully acknowledge the cooperation of Kamioka Mining and Smelting Company.
 This work was supported by the Japanese Ministry of Education, Culture, Sports, Science and Technology, and Grant-in-Aid for Scientific Research (B) 16340066.
 We are supported by Japan Society for the Promotion of Science.

\end{document}